\documentclass[12pt, draftclsnofoot, onecolumn]{IEEEtran}
\usepackage[T1]{fontenc}
\usepackage{rotating,graphics,psfrag,epsfig}
\usepackage{graphicx}

\usepackage{cite}
\usepackage{mathtools}
\usepackage{flushend}
\usepackage{xcolor}
\usepackage{amsmath}
\usepackage{amsfonts}
\usepackage{amssymb}
\usepackage[linesnumbered,ruled]{algorithm2e}
\usepackage{array}
\usepackage{gensymb}
\usepackage[printonlyused,withpage]{acronym}

\newcommand{\qed}{\nobreak \ifvmode \relax \else
      \ifdim\lastskip<1.5em \hskip-\lastskip
      \hskip1.5em plus0em minus0.5em \fi \nobreak
       \vrule height0.75em width0.5em depth0.25em\fi}
\usepackage{url}       
\acrodef{DFT}{discrete Fourier transform}
\acrodef{IFFT}{inverse fast Fourier transform}
\acrodef{FFT}{fast Fourier transform}
\acrodef{SNR}{signal-to-noise ratio}
\acrodef{CP}{cyclic prefix}
\acrodef{TDD}{time division duplex}
\acrodef{FDD}{frequency division duplex}
\acrodef{ISI}{inter symbol interference}
\acrodef{PA}{Power-Amplifier}
\acrodef{AWGN}{additive white Gaussian noise}
\acrodef{i.i.d.}{independent and identically distributed}
\acrodef{mm-wave}{millimeter wave}
\acrodef{cm-wave}{centimeter-wave}
\acrodef{SMD}{sparsity mask detection}
\acrodef{MIMO}{multiple-input-multiple-output}
\acrodef{SNR}{signal-to-noise ratio}
\acrodef{FIM}{Fisher information matrix}
\acrodef{CRB}{Cram\'{e}r-Rao bound}
\acrodef{LLR}{log-likelihood ratio}
\acrodef{MAP}{maximum a posteriori probability}
\acrodef{SBL}{sparse Bayesian learning}
\acrodef{DF}{dynamic filtering}
\acrodef{MSE}{mean square error}
\acrodef{VMF}{Von-Mises-Fisher}
\acrodef{DL}{downlink}
\acrodef{ZF}{zero-forcing}
\acrodef{NF}{normalizing flow}
\acrodef{ELBO}{evidence lower bound}
\acrodef{KL}{Kullback-Leibler}
\acrodef{UL}{uplink}
\acrodef{DL}{downlink}
\acrodef{DF}{dynamic filtering}
\acrodef{IAF}{inverse auto-regressive flow}
\acrodef{GAN}{generative adversarial network}
\acrodef{CWNA}{continuous white noise acceleration}
\acrodef{UWB}{ultra-wide bandwidth}
\acrodef{QAM}{quadrature amplitude modulation}
\acrodef{GPS}{global positioning system}
\acrodef{BF}{beamforming}
\acrodef{i.i.d.}{independent and identically distributed}
\acrodef{QoS}{quality-of-service}
\acrodef{LOS}{line-of-sight}
\acrodef{ROC}{receiver operating characteristic}
\acrodef{CFAR}{constant false alarm rate}
\acrodef{SD}{support detection}
\acrodef{QCD}{quickest change detection}
\acrodef{TCD}{transient change detection}
\acrodef{FMA}{finite moving average}
\acrodef{WLC}{window-limited CUSUM}
\acrodef{CUSUM}{cumulative sum}
\acrodef{NLOS}{non-line-of-sight}
\acrodef{MLP}{multi-layer perceptron}
\acrodef{OLOS}{obstructed-line-of-sight}
\acrodef{RF}{radio-frequency}
\acrodef{EXIP}{extended invariance principle}
\acrodef{DLS}{damped least-squares}
\acrodef{CDF}{cumulative distribution function}
\acrodef{MPCs}{multi-path components}
\acrodef{ML}{maximum likelihood}
\acrodef{MS}{mobile station}
\acrodef{MT}{multi-task}
\acrodef{pdf}{probability density function}
\acrodef{WLS}{weighted least squares}
\acrodef{LMA}{Levenberg-Marquardt algorithm}
\acrodef{GNA}{Gauss-Newton algorithm}
\acrodef{BS}{base station}
\acrodef{ADC}{analog-to-digital-converter}
\acrodef{AOA}{angle-of-arrival}
\acrodef{UL-AOA}{uplink AOA}
\acrodef{DL-AOA}{downlink AOA}
\acrodef{UL-AOD}{uplink AOD}
\acrodef{DL-AOD}{downlink AOD}
\acrodef{RTS}{Rauch-Tung-Striebel}
\acrodef{pdf}{probability density function}
\acrodef{DOA}{direction-of-arrival}
\acrodef{AOD}{angle-of-departure}
\acrodef{TOA}{time-of-arrival}
\acrodef{TDOA}{time-difference-of-arrival}
\acrodef{ULA}{uniform linear array}
\acrodef{PSD}{positive semidefinite}
\acrodef{EFIM}{equivalent Fisher information matrix}
\acrodef{FB}{fractional bandwidth}
\acrodef{REB}{rotation error bound}
\acrodef{PEB}{position error bound}
\acrodef{SDL}{sensor delay line}
\acrodef{TDL}{tapped delay line}
\acrodef{CSI}{channel state information}
\acrodef{OMP}{orthogonal matching pursuit}
\acrodef{DCS-SOMP}{distributed compressed sensing-simultaneous orthogonal matching pursuit}
\acrodef{DCS}{distributed compressed sensing}
\acrodef{SS-UKF}{spherical simplex unscented Kalman filter}
\acrodef{UKF}{unscented Kalman filter}
\acrodef{EKF}{extended Kalman filter}
\acrodef{KF}{Kalman filter}
\acrodef{CS}{compressed sensing}
\acrodef{CoSOMP}{compressive sampling matched pursuit}
\acrodef{SOMP}{simultaneous orthogonal matching pursuit}
\acrodef{RA-ORMP}{rank-aware order recursive matching pursuit}
\acrodef{G-BPDN}{group basis pursuit denoising}
\acrodef{GCS}{group sparse compressed sensing}
\acrodef{MMV}{multiple measurement vectors}
\acrodef{SMV}{single measurement vector}
\acrodef{ReMBo}{reduce MMV and boost}
\acrodef{MLE}{maximum likelihood estimation}
\acrodef{IQML}{iterative quadratic maximum likelihood}
\acrodef{RMSE}{root-mean-square error}
\acrodef{LS}{least squares}
\acrodef{RSE}{root-square error}
\acrodef{rsCRB}{root-square CRB}
\acrodef{RMS}{root-mean-square}
\acrodef{MMSE}{minimum mean square error}
\acrodef{EM}{expectation maximization}
\acrodef{SAGE}{space-alternating generalized expectation maximization}
\acrodef{OFDM}{orthogonal frequency division multiplexing}

\usepackage[caption=false, font=footnotesize]{subfig}
\begin{document} 
Note: This work has been submitted to the IEEE for possible publication. Copyright may be transferred without notice, after which this version may no longer be accessible.
\newpage                       
\title{Sparse Bayesian Multi-Task Learning of Time-Varying Massive MIMO Channels with Dynamic Filtering}
\author{Arash Shahmansoori
\thanks{
Arash Shahmansoori is with the Mitsubishi Electric R\& D Center Europe, Rennes, France, email: arash.mansoori65@gmail.com (Corresponding author: Arash Shahmansoori.)
}
}   
\maketitle
\begin{abstract}
Sparsity of channel in the next generation of wireless communication for massive \ac{MIMO} systems can be exploited to reduce the overhead in the training. The \ac{MT}-\ac{SBL} is applied for learning time-varying sparse channels in the uplink for multi-user massive \ac{MIMO} orthogonal frequency division multiplexing systems. In particular, the dynamic information of the sparse channel is used to initialize the hyperparameters in the \ac{MT}-\ac{SBL} procedure for the next time step. Then, the expectation maximization based updates are applied to estimate the underlying parameters for different subcarriers. Through the simulation studies, it is observed that using the dynamic information from the previous time step considerably reduces the complexity and the required time for the convergence of \ac{MT}-\ac{SBL} algorithm with negligible sacrificing of the estimation accuracy. Finally, the power leakage is reduced due to considering angular refinement in the proposed algorithm.   
\end{abstract}
\section{Introduction}
\IEEEPARstart{M}{assive} \ac{MIMO} systems has become a potential candidate to meet the need for high throughput for the next generation cellular networks \cite{6375940,6736746}. To this end, the \ac{CSI} at the \ac{BS} should be available. The estimation of massive \ac{MIMO} channels without considering the sparsity results unacceptable overhead in both \ac{TDD} and \ac{FDD} modes.

Despite several publications regarding the training of the massive \ac{MIMO} systems for the static channels, there are a few publications for the case of time-varying massive \ac{MIMO} systems, e.g., \cite{6827171,7996914,8410591}. Real time tracking of the beamformer is proposed in \cite{6827171}, channel tracking under time and spatial varying scenarios based on the \ac{EKF} is proposed in \cite{7996914}. In the most recent publication, channel estimation in the \ac{UL}/\ac{DL} is considered in both \ac{TDD} and \ac{FDD} modes by learning the spatial information using the virtual channel representation \cite{8410591}. In this approach, the \ac{SBL} is applied to learn the spatial characteristics of the sparse channel with the \ac{KF} and \ac{RTS} to track the posterior statistics in the expectation step with a searching method in the maximization step. However, power leakage is not compensated that results reducing the estimation accuracy. 

In this letter, we consider a massive \ac{MIMO} \ac{OFDM} system as the \ac{MT}-\ac{SBL} problem and propose a novel dynamic filtering based approach with off-grid refinement. Our contributions are summarized as follows.
\begin{itemize}
\item Inspired by \cite{2019arXiv190205362O}, a novel dynamic filtering for multiple measurement modified sparse channel vector for the dictionary matrix with angular refinement is proposed. Then, it is argued that the angular refinement acts as a spatial rotation operation with first order Taylor series approximation for power leakage reduction.
\item An efficient initialization for the traditional \ac{SBL} problem is obtained to reduce the complexity and the required time for convergence. The proposed initialization gains from the dynamic information of the previous time step. Unlike the \ac{KF} based tracking in the literature that requires the estimation of underlying dynamic model \cite{8410591}, it is assumed that the underlying dynamic model for channel variation is unknown.
\item Simulation results demonstrate that the proposed method outperforms the existing methods from the literature in terms of tracking accuracy and the required number of iterations for convergence.
\end{itemize}
\begin{figure}   
\psfrag{a}[][c]{\footnotesize $\boldsymbol{\alpha}$}
\psfrag{a0}[][c]{\footnotesize $\alpha_{0}$}
\psfrag{h}[][c]{\footnotesize $\mathbf{h}_{\mathrm{BS}}[n]$}
\psfrag{n}[][c]{\footnotesize $\tilde{\mathbf{n}}[n]$}
\psfrag{y}[][c]{\footnotesize $ \color{white}\tilde{\mathbf{y}}^{\mathrm{ul}}[n]$}
\psfrag{ns}[][r]{\footnotesize $n=0,\ldots,N-1$}
\psfrag{t-1}[][]{\footnotesize $t-1$}
\psfrag{t}[][]{\footnotesize $t$}
\psfrag{v}[][]{\footnotesize $\boldsymbol{\nu}$}
\psfrag{del}[][]{\footnotesize $\delta$}
\psfrag{ft}[][]{\footnotesize $\mathbf{f}_{t}(.)$}
\psfrag{c}[][c]{\footnotesize $\mathbf{c}$}
\psfrag{d}[][c]{\footnotesize $\mathbf{d}$}
\psfrag{hh}[][c]{\footnotesize $\hat{\mathbf{h}}^{(t-1)}_{\mathrm{BS}}[n]$}
\psfrag{hhh}[][c]{\footnotesize $\tilde{\mathbf{h}}^{(t)}_{\mathrm{BS}}[n]$}
\psfrag{s}[][c]{\footnotesize $\hat{\mathbf{s}}^{(t-1)}[n]$}
\psfrag{ss}[][c]{\footnotesize $\tilde{\mathbf{s}}^{(t)}[n]$}
\psfrag{hb}[][c]{\footnotesize $\hat{\mathbf{h}}^{(t)}_{\mathrm{BS}}[n]$}
\centering
\includegraphics[width=0.8\columnwidth]{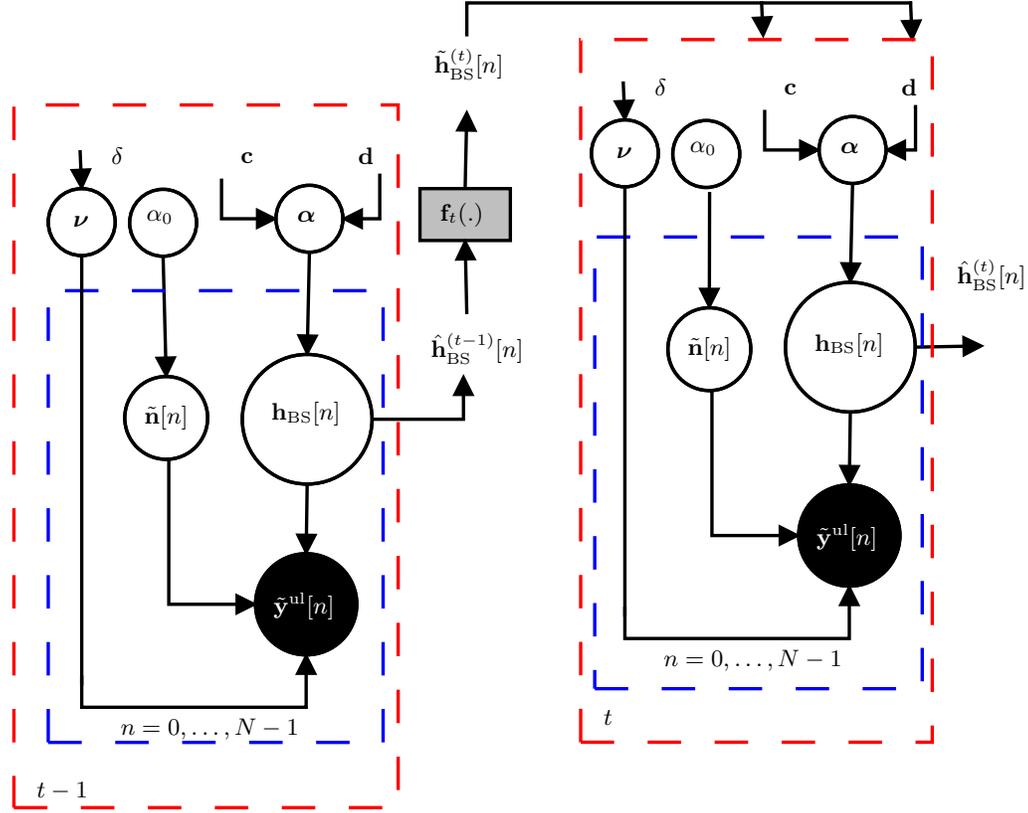}
  \caption{Hierarchical model representation of Bayesian multi-task sparse channel learning with \ac{DF} and the unknown underlying dynamic operation $\mathbf{f}_{t}(.)$.}
  \label{Hierarchical}
\end{figure}
\section{The Proposed Method}
In this section, a method for tracking sparse communication channel based on Bayesian learning with dynamic filtering is proposed. The hierarchical model representation of such a problem is shown in Fig. \ref{Hierarchical}. First, a system model for multi-user \ac{MIMO}-\ac{OFDM} in beamspace is presented in the \ac{UL}. Then, the \ac{MT}-\ac{SBL} for training and tracking with the efficient initialization is proposed with the corresponding algorithm. 
\subsection{System Model}
The sparse system model for multi-user \ac{MIMO}-\ac{OFDM} in beamspace can be approximated by the sparse model in the \ac{UL} as
\begin{equation}\label{eq1}
\tilde{\mathbf{y}}^{\mathrm{ul}}_{t}[n]=\tilde{\mathbf{\Upsilon}}_{\mathrm{BS},\boldsymbol{\nu}}[n]\mathbf{h}^{(t)}_{\mathrm{BS}}[n]+\tilde{\mathbf{n}}_{t}[n],
\end{equation}
where $\tilde{\mathbf{y}}^{\mathrm{ul}}_{t}[n]\in\mathbb{C}^{N_{\mathrm{BS}}L_{m}}$ denotes the received signal with the training sequences of length $L_{m}$ for the users, $m=1,\ldots,M$, and the $N_{\mathrm{BS}}$ received antennas in the \ac{BS}. The term $\mathcal{CN}(\tilde{\mathbf{n}}_{t}[n]\vert\mathbf{0},\alpha^{-1}_{0}\mathbf{I})$ is the \ac{i.i.d.} zero-mean Gaussian measurement noise with unknown precision $\alpha_{0}$ (i.e., variance $1/\alpha_{0}$), and $\mathbf{h}^{(t)}_{\mathrm{BS}}[n]\in\mathbb{C}^{MN_{\mathrm{BS}}}$ denotes the sparse channel vector for $M$ users at $t$-th time step with the same spatial support for subcarriers\footnote{The elements of angle-delay domain channel response are approximately mutually statistically uncorrelated for sufficiently large $N_{\mathrm{BS}}$ and the transmission can be over one or multiple \ac{OFDM} symbols \cite{7332961}.} $n=0,\ldots,N-1$. The sparse channel vector $\mathbf{h}^{(t)}_{\mathrm{BS}}[n]$ is defined as
\begin{equation}\label{eq1a}
\mathbf{h}^{(t)}_{\mathrm{BS}}[n]=\begin{bmatrix}
\left(\mathbf{h}^{(t)}_{\mathrm{BS},1}[n]\right)^{\mathrm{T}}&\ldots &\left(\mathbf{h}^{(t)}_{\mathrm{BS},M}[n]\right)^{\mathrm{T}}
\end{bmatrix}^{\mathrm{T}},
\end{equation}
where\footnote{The extension to the multi-antenna users, and considering the effect of coupling is straightforward.} $\mathbf{h}^{(t)}_{\mathrm{BS},m}[n]$ denotes the modified virtual channel representation for the $m$-th user, over $t$-th time step, and subcarrier $n$ \cite{7524027}. The dictionary $\tilde{\mathbf{\Upsilon}}_{\mathrm{BS},\boldsymbol{\nu}}[n]\in\mathbb{C}^{N_{\mathrm{BS}}L_{m}\times N_{\mathrm{BS}}M}$, is given as 
\begin{equation}\label{eq1b}
\tilde{\mathbf{\Upsilon}}_{\mathrm{BS},\boldsymbol{\nu}}[n]=\begin{bmatrix}
\mathbf{\Upsilon}^{(1)}_{\mathrm{BS},\boldsymbol{\nu}}[n]&\ldots &\mathbf{\Upsilon}^{(M)}_{\mathrm{BS},\boldsymbol{\nu}}[n]
\end{bmatrix},
\end{equation}
where
\begin{align}
\mathbf{\Upsilon}^{(m)}_{\mathrm{BS},\boldsymbol{\nu}}[n]&=\mathbf{x}_{m}[n]\otimes \mathbf{\Omega}_{\mathrm{BS}}(\boldsymbol{\nu}),\label{eq2}\\
\mathbf{\Omega}_{\mathrm{BS}}(\boldsymbol{\nu})&=\bar{\mathbf{F}}_{\mathrm{BS}}+\dot{\bar{\mathbf{F}}}_{\mathrm{BS}}\mathrm{diag}\{\boldsymbol{\nu}\},\label{eq3}
\end{align}
in which $\mathbf{x}_{m}[n]\in\mathbb{C}^{L_{m}}$. The matrix $\bar{\mathbf{F}}_{\mathrm{BS}}\in\mathbb{C}^{N_{\mathrm{BS}}\times N_{\mathrm{BS}}}$ is the normalized \ac{DFT} matrix. The term $\dot{\bar{\mathbf{F}}}_{\mathrm{BS}}$ denotes the first order \ac{DFT} based differentiation, and $\boldsymbol{\nu}\in\mathbb{R}^{N_{\mathrm{BS}}}$ is the off-grid vector. The expression \eqref{eq3} can be interpreted as spatial rotation operation with first order Taylor series approximation to mitigate leakage and strengthen channel sparsity \cite{7524027}. It is assumed that the underlying dynamic function $\mathbf{f}_{t}(.)$ is unknown. Consequently, a Gaussian blurring kernel for the spatial domain on the previous time step $t-1$ with the error in the dynamic model $\mathcal{N}(\mathbf{w}_{t}\vert\mathbf{0},\mathbf{Q})$ is used as the dynamic equation\footnote{It is worth noting that the proposed method can be reformulated based on the estimated values for the $T_{b}>1$ blocks instead of $T_{b}=1$ block. In other words, the smoothed version of the proposed dynamic filtering method is simply achieved by a chain of $T_{b}$ successive time steps and estimation of the corresponding sparse vectors starting from the initial estimation of the sparse vector for the time step zero obtained by the \ac{SBL} method. This is essentially similar to the \ac{RTS} smoother.}.
\subsection{\ac{MT}-\ac{SBL} learning method with \ac{DF}}
Considering the measurement model \eqref{eq1}, the likelihood function is obtained as\footnote{To simplify the notation, the subscript $t$ is dropped in this subsection.}
\begin{equation}\label{eq7}
p(\tilde{\mathbf{y}}^{\mathrm{ul}}[n]\vert\mathbf{h}_{\mathrm{BS}}[n],\boldsymbol{\alpha},\alpha_{0},\boldsymbol{\nu}) = (\pi/\alpha_{0})^{-L_{m}N_{\mathrm{BS}}}\mathrm{exp}\left(-\alpha_{0}\Vert\tilde{\mathbf{y}}^{\mathrm{ul}}[n]-\tilde{\mathbf{\Upsilon}}_{\mathrm{BS},\boldsymbol{\nu}}[n]\mathbf{h}_{\mathrm{BS}}[n]\Vert_{2}^{2}\right).
\end{equation}
A zero-mean Gaussian prior is placed on $\mathbf{h}_{\mathrm{BS}}[n]$
\begin{equation}\label{eq8}
p(\mathbf{h}_{\mathrm{BS}}[n]\vert\boldsymbol{\alpha})=\prod_{l=1}^{MN_{\mathrm{BS}}}\mathcal{CN}\left(h_{l,\mathrm{BS}}[n]\vert 0,\alpha^{-1}_{l}\right),
\end{equation}
with the hyperparameters $\boldsymbol{\alpha}=[\alpha_{1},\ldots,\alpha_{MN_{\mathrm{BS}}}]^{\mathrm{T}}$ shared among all the $N$ tasks. The term $h_{l,\mathrm{BS}}[n]$ denotes the $l$-th element of $\mathbf{h}_{\mathrm{BS}}[n]$ for $l=1,\ldots,MN_{\mathrm{BS}}$. Gamma priors are placed on the parameters $\alpha_{l}$ and $\alpha_{0}$ to complete the probability model as
$p(\alpha_{l}\vert c_{l},d_{l})=\mathcal{G}\left(\alpha_{l}\vert c_{l},d_{l}\right)$, and
$p(\alpha_{0}\vert a,b)=\mathcal{G}\left(\alpha_{0}\vert a,b\right)$
where $\{c_{l}\},a$ and $\{d_{l}\},b$ denote the shape and scale parameters in the Gamma distributions, respectively. Finally, the off-grid parameter is obtained by a uniform prior distribution as 
$p(\nu_{r}\vert \delta)=\mathcal{U}_{\nu_{r}}\left(\left[-\frac{1}{2}\delta, \frac{1}{2}\delta\right]\right)$
where $\mathcal{U}_{\nu_{r}}([-\delta/2, \delta/2])$ for $r=1,\ldots,N_{\mathrm{BS}}$ denotes a uniform distribution with the grid size $\delta$ and probability $1/\delta$ in the interval $[-\delta/2, \delta/2]$ and zero elsewhere. The value of $\delta$ is set to $2\pi/N_{\mathrm{BS}}$ as $\nu_{r}$ acts as a spatial rotation parameter for $N_{\mathrm{BS}}$-point \ac{DFT}. The full posterior distribution is decomposed as
\begin{equation}\label{eq11}
p(\mathbf{h}_{\mathrm{BS}}[n],\boldsymbol{\alpha},\alpha_{0},\boldsymbol{\nu}\vert\tilde{\mathbf{y}}^{\mathrm{ul}}[n])=p(\mathbf{h}_{\mathrm{BS}}[n]\vert\tilde{\mathbf{y}}^{\mathrm{ul}}[n],\boldsymbol{\alpha},\alpha_{0},\boldsymbol{\nu})\\\times\: p(\boldsymbol{\alpha},\alpha_{0},\boldsymbol{\nu}\vert\tilde{\mathbf{y}}^{\mathrm{ul}}[n]),
\end{equation}
where the posterior on the channel vector is $p(\mathbf{h}_{\mathrm{BS}}[n]\vert\tilde{\mathbf{y}}^{\mathrm{ul}}[n],\boldsymbol{\alpha},\alpha_{0},\boldsymbol{\nu})=\mathcal{CN}(\mathbf{h}_{\mathrm{BS}}[n]\vert\boldsymbol{\mu}[n],\mathbf{\Sigma}[n])$ with
\begin{align}
\boldsymbol{\mu}[n]&=\alpha_{0}\mathbf{\Sigma}[n]\tilde{\mathbf{\Upsilon}}^{\mathrm{H}}_{\mathrm{BS},\boldsymbol{\nu}}[n]\tilde{\mathbf{y}}^{\mathrm{ul}}[n],\label{eq12}\\
\mathbf{\Sigma}[n]&=\left(\mathrm{diag}\left\{\boldsymbol{\alpha}\right\}+\alpha_{0}\tilde{\mathbf{\Upsilon}}^{\mathrm{H}}_{\mathrm{BS},\boldsymbol{\nu}}[n]\tilde{\mathbf{\Upsilon}}_{\mathrm{BS},\boldsymbol{\nu}}[n]\right)^{-1}.\label{eq13}
\end{align}
Assuming uninformative prior on the noise precision $\alpha_{0}$, i.e., $a=b=0$, the hyperparameter $\boldsymbol{\alpha}$ is obtained by the \ac{MAP} estimation based on the marginal log likelihood $\mathcal{L}(\boldsymbol{\alpha},\alpha_{0})$ defined as
\begin{equation}\label{eq15}
\mathcal{L}(\boldsymbol{\alpha},\alpha_{0})\propto \sum_{n=0}^{N-1}\log\vert\mathbf{C}[n]\vert+\left(\tilde{\mathbf{y}}^{\mathrm{ul}}[n]\right)^{\mathrm{H}}\mathbf{C}^{-1}[n]\tilde{\mathbf{y}}^{\mathrm{ul}}[n]+2N\sum_{l=1}^{MN_{\mathrm{BS}}}\left(c_{l}\log\alpha_{l}-d_{l}\alpha_{l}\right),
\end{equation}
where 
$\mathbf{C}[n]=\alpha^{-1}_{0}\mathbf{I}+\tilde{\mathbf{\Upsilon}}_{\mathrm{BS},\boldsymbol{\nu}}[n]\left(\mathrm{diag}\left\{\boldsymbol{\alpha}\right\}\right)^{-1}\tilde{\mathbf{\Upsilon}}^{\mathrm{H}}_{\mathrm{BS},\boldsymbol{\nu}}[n]$.
Next, the hyperparameters $\{c_{l},d_{l}\}$ are obtained to minimize the \ac{MSE} between the dynamic prediction of sparse channel vector from the previous time step denoted as $\pmb{\hbar}_{\mathrm{BS}}[n]$, and the \ac{SBL} with \ac{DF} in \eqref{eq12} as
\begin{equation}\label{eq17}
\boldsymbol{\alpha}_{\mathrm{opt}}=\underset{\boldsymbol{\alpha}}{\mathrm{argmin}}\:\sum_{n=0}^{N-1}
\mathbb{E}\left[\|\boldsymbol{\mu}[n]-\pmb{\hbar}_{\mathrm{BS}}[n]\|^{2}_{2}\right].
\end{equation}
Taking the derivative of \eqref{eq17} with respect to $\boldsymbol{\alpha}$ and setting the result to zero, and assuming equal transmit power for different subcarriers without loss of generality, $\Vert \mathbf{x}_{m}[n]\Vert_{2}^2=P_{m},\: n=0,\ldots,N-1, m=1,\ldots,M$, leads to (see Appendix \ref{app1})
\begin{equation}\label{eq18}
\alpha_{\mathrm{opt},l}=\frac{1}{\frac{1}{N}\sum_{n=0}^{N-1}\vert \hbar_{l,\mathrm{BS}}[n]\vert^{2}},\: l=1,\ldots,MN_{\mathrm{BS}},
\end{equation}
where $\hbar_{l,\mathrm{BS}}[n]=\left[\pmb{\hbar}_{\mathrm{BS}}[n]\right]_{l}$.
Considering the fact that $\alpha_{l}=c_{l}/d_{l}$ minimizes the second term in \eqref{eq15}, and comparing with \eqref{eq18}, the parameters $c_{l}$ and $d_{l}$ can be obtained as\footnote{It is worth noting that there are other options such that the identity $\alpha_{l}=c_{l}/d_{l}$ is fulfilled. However, we found that using these values result a robust performance in the tracking phase.} 
\begin{equation}
c_{l} = \alpha_{\mathrm{opt},l},\: d_{l} = 1 \label{eq19}.
\end{equation}
\subsubsection*{Practical note}For the case that the value of $\alpha_{\mathrm{opt},l}$ is large, i.e., $\frac{1}{N}\sum_{n=0}^{N-1}\vert \hbar_{l,\mathrm{BS}}[n]\vert^{2}$ is small, and consequently Gamma distribution has a large shape value $c_{l}$, we set $c_{l}=\sqrt{\alpha_{\mathrm{opt},l}}$ to avoid sticking in a local minimum and provide more robust performance and numerical stability.

For the case that $c_{l}=d_{l}=0$, there is no dynamic information from the previous time step. Consequently, the estimation of corresponding parameters is obtained under no prior information. It is worth noting that user grouping can be performed after this stage such that users in the same group does not share the same spatial information and orthogonal pilots are applied in different groups \cite{8410591}. In this case, the optimal values of $\boldsymbol{\alpha}$, $\alpha_{0}$, and $\boldsymbol{\nu}$ can be obtained by maximizing the expectation of the joint probability $p(\mathbf{h}_{\mathrm{BS}}[n],\boldsymbol{\alpha},\boldsymbol{\nu},\alpha_{0},\tilde{\mathbf{y}}^{\mathrm{ul}}[n])$ with respect to the posterior distribution on $\mathbf{h}_{\mathrm{BS}}[n]$. The results of this maximization for $\boldsymbol{\alpha}$ and $\alpha_{0}$ are summarized as follows. First, the parameters $\alpha_{l}$ and $\alpha_{0}$ are obtained by \ac{EM} procedure with iterative updates by treating $\mathbf{h}_{\mathrm{BS}}[n]$ as the hidden variable as
\begin{align}
\alpha_{l} &= \frac{c_{l}-1+N}{d_{l}+\sum_{n=0}^{N-1}\left[\mathbf{\Sigma}[n]\right]_{l,l}+\sum_{n=0}^{N-1}\vert\mu_{l}[n]\vert^{2}},\label{eq21}\\
\alpha_{0} &= \frac{N_{\mathrm{BS}}L_{m}N+a-1}{\sum_{n=0}^{N-1}T^{(a)}_{\boldsymbol{\nu}}[n]+\sum_{n=0}^{N-1}T^{(b)}_{\boldsymbol{\nu}}[n]+b},\label{eq22}
\end{align}
where
\begin{align}
T^{(a)}_{\boldsymbol{\nu}}[n] &= \Vert\tilde{\mathbf{y}}^{\mathrm{ul}}[n]-\tilde{\mathbf{\Upsilon}}_{\mathrm{BS},\boldsymbol{\nu}}[n]\mathbf{h}_{\mathrm{BS}}[n]\Vert_{2}^{2},\label{eqnew1a}\\
T^{(b)}_{\boldsymbol{\nu}}[n] &= \mathrm{tr}\left\{\tilde{\mathbf{\Upsilon}}^{\mathrm{H}}_{\mathrm{BS},\boldsymbol{\nu}}[n]\tilde{\mathbf{\Upsilon}}_{\mathrm{BS},\boldsymbol{\nu}}[n]\mathbf{\Sigma}[n]\right\}.\label{eqnew1b}
\end{align}
The off-grid estimation of \ac{AOA} is obtained by solving the linear equation for $\boldsymbol{\nu}$ with taking the derivative of the term
\begin{equation}\label{eq2_app3}
\sum_{n=0}^{N-1}T^{(a)}_{\boldsymbol{\nu}}[n]+T^{(b)}_{\boldsymbol{\nu}}[n],
\end{equation}
\begin{algorithm}\label{alg_sbl}
\SetKwInOut{Input}{Input}
\SetKwInOut{Output}{Output}
\Input{Received signal $\tilde{\mathbf{y}}^{\mathrm{ul}}_{t}[n]$, dictionary matrix $\bar{\mathbf{F}}_{\mathrm{BS}}$ and the first order derivative of dictionary matrix $\dot{\bar{\mathbf{F}}}_{\mathrm{BS}}$, the number of subcarriers $N$, the number of iterations $I_{\mathrm{iter}}$, the number of users $M$, and stop threshold $\beta_{\mathrm{th}}$.}
\Output{Estimated $\hat{\mathbf{h}}^{(t)}_{\mathrm{BS}}[n]$ for $\forall n$ and $\forall t$.}
\textit{Initialisation}: Set $\alpha_{0}=1$ and the hyperparameters $a=b=0.01$, $\boldsymbol{\alpha}=\mathbf{1}_{N_\mathrm{BS}}$ and the hyperparameters $c_{l}=d_{l}=0.01$ for $l=1,\ldots,MN_{\mathrm{BS}}$, and $\boldsymbol{\nu}=\mathbf{0}$.

\SetAlgoLined
\DontPrintSemicolon
\For {$t = 1$ to $T+1$}{
\While{$\rho\leq\beta_{\mathrm{th}}$ or $i_{\mathrm{iter}}\leq I_{\mathrm{iter}}$} {
Form $\tilde{\mathbf{\Upsilon}}_{\mathrm{BS},\boldsymbol{\nu}}[n]$ based on \eqref{eq1b}.\;
Compute $\boldsymbol{\mu}[n]$ from \eqref{eq12}, and $\mathbf{\Sigma}[n]$ from \eqref{eq13}.\;
Set $\boldsymbol{\alpha}'=\boldsymbol{\alpha}$, and update $\boldsymbol{\alpha}$ from \eqref{eq21} for time step $t-1$.\;
Update $\alpha_{0}$ from \eqref{eq22} for time step $t-1$.\;
Update $\boldsymbol{\nu}$ based on \eqref{eq2_app3} for time step $t-1$.\;
If $i_{\mathrm{iter}}>1$, then $\rho=\Vert\boldsymbol{\alpha}-\boldsymbol{\alpha}'\Vert_{2}/\Vert\boldsymbol{\alpha}'\Vert_{2}$.\;
}
Compute $\pmb{\hbar}_{\mathrm{BS}}[n]=\hat{\mathbf{h}}^{(t-1)}_{\mathrm{BS}}[n]$.\;
For the time step $t$, calculate $c_{l}$ and $d_{l}$ from \eqref{eq19}.\;
}
\caption{\ac{MT}-\ac{SBL} with \ac{DF} in the \ac{UL}}
\end{algorithm}
with respect to $\boldsymbol{\nu}$ and setting the result to zero.
\subsection{The Algorithm}
Algorithm \ref{alg_sbl} summarizes the proposed method for \ac{MT}-\ac{SBL} learning in the \ac{UL} with dynamic filtering. Starting from the input parameters and initialization assuming no dynamic information at the beginning, i.e., $c_{l}=d_{l}=0.01$, the proposed method iterates for $T$ time steps and updates the underlying parameters using the dynamic information provided by the previous time step based on \eqref{eq19} with considering the practical note. It will be shown in the simulation results that using the dynamic information considerably reduces the required number of iterations for convergence compared to the traditional \ac{SBL} method.

It is worth noting that the proposed method can be easily adopted for the \ac{DL} tracking in both \ac{FDD} or \ac{TDD} mode. For the \ac{TDD} mode, the underlying parameters are exactly the same due to channel reciprocity, while in the \ac{FDD} mode the method is adjusted for the array response with the new wavelength $\tilde{\lambda}_{c}$.
\section{Simulation Results}
The performance of the proposed method is numerically evaluated in this section. A massive \ac{MIMO} system with $N_{\mathrm{BS}}=64$ antenna elements with half wavelength spacing in the \ac{BS} is considered. The \ac{AOA} of each user is randomly selected from\footnote{The performance for the users with different spatial supports in the training and tracking phases is investigated.} $[-80\degree, 80\degree]$ with the angular spread on the order of $2\degree$ for $M=2$, $\beta_{\mathrm{th}}=1e-3$, $I_{\mathrm{iter}}=1e3$, $N=40$, and $\mathrm{SNR}\:[\mathrm{dB}]=10$. The initialization of the hyper-parameters is achieved with the values as in the algorithm \ref{alg_sbl}. The results are averaged over $100$ realizations. Finally, the maximum angular variations between consecutive time steps for a given environment is on the order of $0.5\degree$.

\begin{figure}[t]   
\psfrag{Average Required Number of Iterations}[][c]{\footnotesize Required number of iterations}
\psfrag{Method in fefei}[][c]{\footnotesize \: Method in \cite{8410591}}
\psfrag{Proposed method}[][c]{\footnotesize \: Proposed method}
\psfrag{True channel}[][c]{\footnotesize \: True channel}
\psfrag{Norm of the Virtual Channel}[][c]{\footnotesize Norm of the virtual channel}
\psfrag{t}[][c]{\footnotesize $t$}
\psfrag{RMSE}[][c]{\footnotesize RMSE}
\centering
\includegraphics[width=0.8\columnwidth]{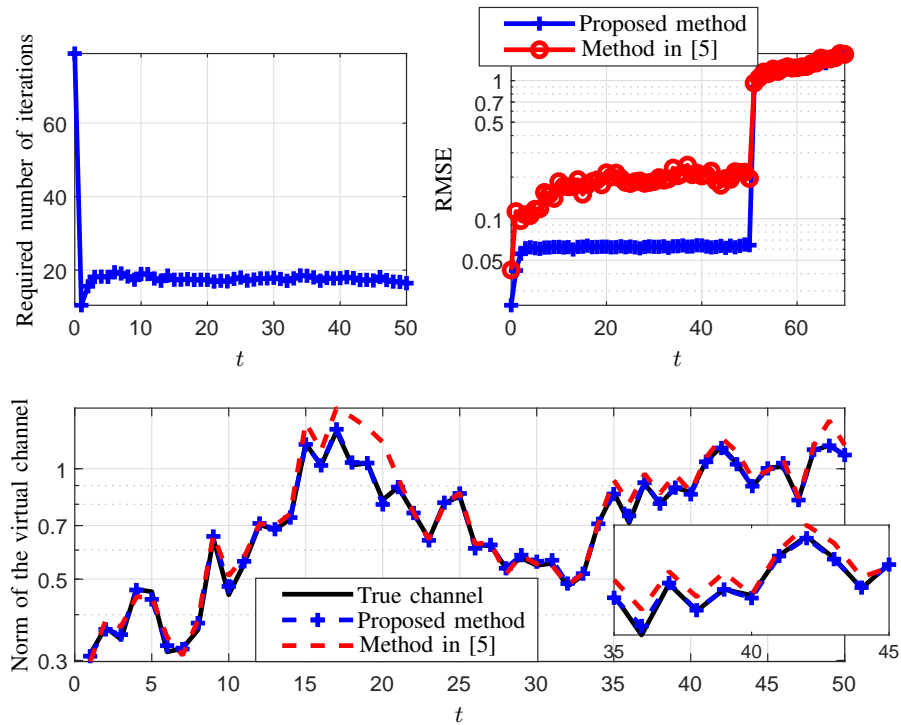}
  \caption{(top-left) Required number of the iterations for the convergence of the algorithm \ref{alg_sbl} with respect to the time steps $t$ averaged over $100$ realizations. (top-right) The \ac{RMSE} of estimated and tracked values of the norm of the virtual channel averaged over subcarriers with respect to the time steps $t$. (bottom) An example of virtual channel tracking over $0<t\leq T$.}
  \label{Sima}
\end{figure}

Fig. \ref{Sima} top left plot shows the required number of iterations for the convergence of the proposed method for different time steps. It is observed that without using the dynamic information at $t=0$ \cite{8537983}, the required number of iterations for convergence of the algorithm \ref{alg_sbl} is considerably more than the case with the prior information provided by the dynamic information at $t>0$. More specifically, the required number of iterations for the convergence of the \ac{MT}-\ac{SBL} method is reduced by around $77.81\%$ compared to the method in \cite{8537983}.

Fig. \ref{Sima} top right plot shows the \ac{RMSE} of the sparse channel over time. It is observed that the proposed method is very effective for virtual channel tracking at $0<t\leq T$ with $T=50$. Moreover, the proposed method provides comparable accuracy to the estimated values at $t=0$ with reduced complexity and required number of iterations. Using the proposed approach, estimation and tracking accuracy are improved by updating the off-grid refinement as a form of spatial rotation operation to mitigate power leakage. This results improved accuracy compared to the adopted method from the literature by around $65.89\%$ \cite{8410591}. An example of virtual channel tracking with improved accuracy for $0<t\leq T$ is provided in Fig. \ref{Sima} bottom plot. Finally, for $t>T$ the users move to a different environment that requires \emph{re-learning} the channel. In other words, using the dynamic information from the previous time step is not sufficient for efficient channel tracking in the new environment. 
\section{Conclusion}
We have studied the tracking of the sparse multi-user massive \ac{MIMO} channels through \ac{MT}-\ac{SBL} learning with dynamic filtering. The proposed method gains the dynamic information and significantly reduces the complexity and required number of iterations for the convergence of estimated parameters. It is worth noting that the proposed method does not assume known underlying dynamic function as apposed to existing methods in the literature based on \ac{KF}. Therefore, the estimation of the underlying dynamic function parameters is not necessary in the proposed method. Through the simulation studies, we demonstrated that the proposed method outperforms the existing approaches in terms of required number of iterations and accuracy. Finally, considering the spatial-wideband effect for time-varying massive \ac{MIMO} systems is an interesting topic for future research \cite{8714079}.
\appendices
\section{Optimal $\boldsymbol{\alpha}$ based on Dynamic Information}\label{app1}
Replacing $\boldsymbol{\mu}[n]$ and $\mathbf{\Sigma}[n]$ from \eqref{eq12} and \eqref{eq13}, respectively, in \eqref{eq17} and the eigenvalue decomposition $\mathbf{U}^{\mathrm{H}}_{\mathrm{BS}}\mathbf{\Lambda}\mathbf{U}_{\mathrm{BS}}=\tilde{\mathbf{\Upsilon}}^{\mathrm{H}}_{\mathrm{BS},\boldsymbol{\nu}}[n]\tilde{\mathbf{\Upsilon}}_{\mathrm{BS},\boldsymbol{\nu}}[n]$, results
\begin{equation}\label{eq1_app1}
\sum_{n=0}^{N-1}\mathbb{E}\left[\|\boldsymbol{\mu}[n]-\pmb{\hbar}_{\mathrm{BS}}[n]\|^{2}_{2}\right]=\sum_{n=0}^{N-1}\sum_{l=1}^{MN_{\mathrm{BS}}}\frac{\alpha^{-1}_{0}\lambda_{l}}{\left(\alpha^{-1}_{0}\alpha_{l}+\lambda_{l}\right)^{2}}+\left(\frac{\lambda_{l}}{\alpha^{-1}_{0}\alpha_{l}+\lambda_{l}}-1\right)^{2}\hbar^{2}_{l}[n],
\end{equation} 
where $\lambda_{l}$ is the $l$-th entry of $\mathbf{\Lambda}$. Equal transmit power for different subcarriers is assumed that results frequency independent eigenvalues $\lambda_{l}$. In computing \eqref{eq1_app1}, we used the fact that the inner product of dictionary matrix can be written as $\tilde{\mathbf{\Upsilon}}^{\mathrm{H}}_{\mathrm{BS},\boldsymbol{\nu}}[n]\tilde{\mathbf{\Upsilon}}_{\mathrm{BS},\boldsymbol{\nu}}[n]=\mathbf{P}\otimes\left(\mathbf{\Omega}^{\mathrm{H}}_{\mathrm{BS}}(\boldsymbol{\nu})\mathbf{\Omega}_{\mathrm{BS}}(\boldsymbol{\nu})\right)$ where $\mathbf{P}=\mathrm{diag}\{\mathbf{p}\}$ is an $M\times M$ diagonal power matrix with $m$-th entry of $P_{m}$ for $m=1,\ldots,M$. Consequently, the eigenvectors of $\tilde{\mathbf{\Upsilon}}^{\mathrm{H}}_{\mathrm{BS},\boldsymbol{\nu}}[n]\tilde{\mathbf{\Upsilon}}_{\mathrm{BS},\boldsymbol{\nu}}[n]$ are independent of signal power $P_{m}$ and off-grid vector $\boldsymbol{\nu}$. The eigenvectors are obtained as $\mathbf{e}_{m}\otimes \mathbf{u}_{r}$ where $\mathbf{e}_{m}$ and $\mathbf{u}_{r}$ denote the $m$-th and $r$-th eigenvectors of $\mathbf{P}$ and $\left(\mathbf{\Omega}^{\mathrm{H}}_{\mathrm{BS}}(\boldsymbol{\nu})\mathbf{\Omega}_{\mathrm{BS}}(\boldsymbol{\nu})\right)$ for $m=1,\ldots,M$ and $r=1,\ldots,N_{\mathrm{BS}}$, respectively. Similarly, the eigenvalues of $\tilde{\mathbf{\Upsilon}}^{\mathrm{H}}_{\mathrm{BS},\boldsymbol{\nu}}[n]\tilde{\mathbf{\Upsilon}}_{\mathrm{BS},\boldsymbol{\nu}}[n]$ are obtained as $\boldsymbol{\lambda}=\mathbf{p}\otimes\boldsymbol{\varsigma}$ where $\boldsymbol{\lambda}$ denotes an $MN_{\mathrm{BS}}\times 1$ vector with the $l$-th entry of $\lambda_{l}$ in which $\mathbf{p}$ and $\boldsymbol{\varsigma}$ are $M\times 1$  and $N_{\mathrm{BS}}\times 1$ vectors denoting the corresponding eigenvalues of $\mathbf{P}$ and $\left(\mathbf{\Omega}^{\mathrm{H}}_{\mathrm{BS}}(\boldsymbol{\nu})\mathbf{\Omega}_{\mathrm{BS}}(\boldsymbol{\nu})\right)$, respectively. Finally, taking the derivative of \eqref{eq1_app1} with respect to $\alpha_{l}$ and setting the result to zero leads to \eqref{eq18}.
\bibliographystyle{IEEEtran}
\bibliography{Final_References}
\end{document}